\documentclass[letter]{aa}
\usepackage{graphicx}
\usepackage{setspace}
\usepackage{natbib}
\usepackage{longtable}
\usepackage{amssymb}
\usepackage[varg]{txfonts}
\bibpunct{(}{)}{;}{a}{}{,}

\begin{document}

\newcommand{\FeII}{[\ion{Fe}{ii}]}
\newcommand{\TiII}{[\ion{Ti}{ii}]}
\newcommand{\SII}{[\ion{S}{ii}]}
\newcommand{\OI}{[\ion{O}{i}]}
\newcommand{\OIp}{\ion{O}{i}}
\newcommand{\PII}{[\ion{P}{ii}]}
\newcommand{\NI}{[\ion{N}{i}]}
\newcommand{\NII}{[\ion{N}{ii}]}
\newcommand{\NIp}{\ion{N}{i}}
\newcommand{\NiII}{[\ion{Ni}{ii}]}
\newcommand{\CaIIp}{\ion{Ca}{ii}}
\newcommand{\PI}{[\ion{P}{i}]}
\newcommand{\CIp}{\ion{C}{i}}
\newcommand{\HeI}{\ion{He}{i}}
\newcommand{\MgIp}{\ion{Mg}{i}}
\newcommand{\MgIIp}{\ion{Mg}{ii}}
\newcommand{\NaI}{\ion{Na}{i}}
\newcommand{\HI}{\ion{H}{i}}
\newcommand{\brg}{Br\,$\gamma$}
\newcommand{\pab}{Pa\,$\beta$}

\newcommand{\macc}{$\dot{M}_{acc}$}
\newcommand{\lacc}{L$_{acc}$}
\newcommand{\lbol}{L$_{bol}$}
\newcommand{\mjet}{$\dot{M}_{jet}$}
\newcommand{\mh}{$\dot{M}_{H_2}$}
\newcommand{\Ne}{n$_e$}
\newcommand{\h}{H$_2$}
\newcommand{\kms}{km\,s$^{-1}$}
\newcommand{\um}{$\mu$m}
\newcommand{\lam}{$\lambda$}
\newcommand{\msyr}{M$_{\odot}$\,yr$^{-1}$}
\newcommand{\Av}{A$_V$}
\newcommand{\msun}{M$_{\odot}$}
\newcommand{\lsun}{L$_{\odot}$}
\newcommand{\cm}{cm$^{-3}$}
\newcommand{\ergscm}{erg\,s$^{-1}$\,cm$^{-2}$}

\newcommand{\bet}{$\beta$}
\newcommand{\alfa}{$\alpha$}

\hyphenation{mo-le-cu-lar pre-vious e-vi-den-ce di-ffe-rent pa-ra-me-ters ex-ten-ding a-vai-la-ble excited}

\title{Spatially resolved \h\ emission from a very low-mass star\thanks{Based on observations collected at the European Southern 
Observatory Paranal, Chile (ESO programme 385.C-0893(A)). The reduced datacube is available in electronic form
at the CDS via anonymous ftp to cdsarc.u-strasbg.fr (130.79.128.5)
or via http://cdsweb.u-strasbg.fr/cgi-bin/qcat?J/A+A/ as a FITS file at the CDS}}
\author{R. Garcia Lopez \inst{1} \and A. Caratti o Garatti \inst{1} \and G. Weigelt \inst{1} \and B. Nisini \inst{2} \and S. Antoniucci \inst{2}}

\offprints{rgarcia@mpifr-bonn.mpg.de}

\institute{Max-Planck-Institut f\"{u}r Radioastronomie, Auf dem H\"{u}gel 69, D-53121 Bonn, Germany \and INAF-Osservatorio Astronomico di Roma via Frascati 33, I-00040, Monteporzio Catone, Italy}

%----------------------------------------------------------------------
%
\date{Received date; Accepted date}
%
%----------------------------------------------------------------------
%
%
\titlerunning{Spatially resolved \h\ 1-0S(1) emission from IRS54}
\authorrunning{Garcia Lopez, R. et al.}

\abstract
{Molecular outflows from very low-mass stars (VLMSs) and brown dwarfs have been studied very little. So far, only a few CO outflows have been observed, allowing us to map the immediate circumstellar environment.}
{We present the first spatially resolved \h\ emission around IRS54 (YLW\,52), a $\sim$0.1-0.2\,\msun\ Class I source. }
{By means of VLT SINFONI K-band observations, we probed the \h\ emission down to the first $\sim$50\,AU from the source.}
{The molecular emission shows a complex structure delineating a large outflow cavity and an asymmetric molecular jet. Thanks to the detection of several \h\ transitions, we are able to estimate average values along the jet-like structure (from source position to knot D) of A$_V\sim$28\,mag, T $\sim$2000-3000\,K, and \h\ column density N(\h)$\sim$1.7$\times$10$^{17}$\,cm$^{-2}$.
%visual extinction of $\sim$28\,mag, temperatures of $\sim$2000-3000\,K and \h\ column density along the outflow of 1.7$\times$10$^{17}$\,cm$^{-2}$. 
This allows us to estimate a mass loss rate of $\sim$2$\times$10$^{-10}$\,\msyr\ for the warm \h\ component .
In addition, from the total flux of the \brg\ line, we infer an accretion luminosity and mass accretion rate of 0.64\,\lsun\ and $\sim$3$\times$10$^{-7}$\,\msyr, respectively. The outflow structure is similar to those found in low-mass Class I and CTTS. However, the \lacc/\lbol\ ratio is very high ($\sim$80\%), and the mass accretion rate is about one order of magnitude higher when compared to objects of roughly the same mass, pointing to the young nature of the investigated source.}
{}


\maketitle

%
%------------------------------------------------------------------------- 
%---------------------------------------------------------------------------

\section{Introduction}
%%%%%%%%%%%%%%%%%%%%%%%%%%%%%%	velocity panels  %%%%%%%%%%%%%%%%%%%%%%%%%%%%%%%%%%%%%%%%%%%%%%%%%%%%%%%%%%%%%%%%%%%%%%%%%%%
\begin{figure}[!th]
\centering
\resizebox{0.76\columnwidth}{!}{\includegraphics{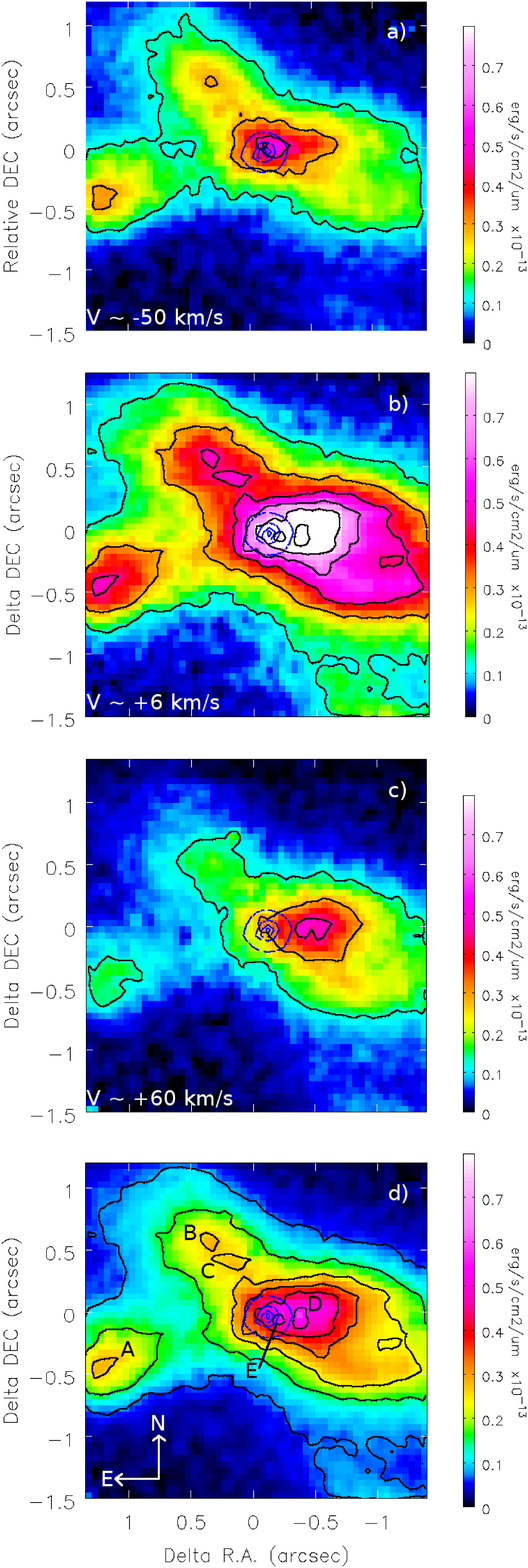}}
\caption{Continuum-subtracted H$_2$\,1-0S(1) images of IRS54. % along five different spectral channels. 
Panels \textit{a} and \textit{c}: average over two spectral channels at $-64$\,\kms\ and -30\,\kms\ (panel a), and +40\,\kms\ and +75\,\kms\ (panel \textit{c}). Panel \textit{b}: Single spectral channel at +6\,\kms. 
The velocities are corrected from an average cloud velocity of 3.5\,\kms\ \citep{wouterloot05,andre07}. 
Panel \textbf{d}: average \h\,1-0S(1) emission over the previous velocity channels. Overplotted are the locations of six spatial regions where spectra were extracted. For reference, contours of the continuum (near the \h\,2.122\,\um\ line) down to the FWHM size 
have been overplotted at the centre of every image (dashed-blue contours). Contours in all panels show values of 1.3, 2.9, 4.5, 5.3, 6.1, 7.7, 9.3$\times$10$^{-14}$\,\ergscm\,\um$^{-1}$.}
\label{fig:vel_channel}
\end{figure}
%%%%%%%%%%%%%%%%%%%%%%%%%%%%%%%%%%%%%%%%%%%%%%%%%%%%%%%%%%%%%%%%%%%%%%%%%%%%%%%%%%%%%%%%%%%%%%%%%%%%%%%%%

Protostellar jets and outflows are associated with the first stages of stellar evolution. They are usually found in active young stellar objects (YSOs), in which a significant fraction of the circumstellar material is still accreting onto the protostar. Jets and outflows can therefore be considered an outcome of the accretion activity.
During the early evolutionary phase, outflows are usually traced through molecular emission lines. Molecular outflows from low- and high-mass protostars have been extensively studied through observations of CO lines \citep{bachiller99,arce07}. These observations have revealed many common observational characteristics between both mass regimes, which points to a common outflow mechanism working from low- to high-mass protostars \citep{arce07}.

Molecular outflows from VLMSs and brown-dwarfs (BDs) have been studied very little, especially during their early evolutionary phase. Only recently have CO submillimetre observations been able to provide direct imaging of a small number of molecular outflows from VLMSs \citep{phan08,phan11}. Most of the few observations of jets from BDs and VLMSs, however, involve relatively evolved, classical TTauri-like YSOs, mainly studied through detecting forbidden emission lines (FELs) \citep{fernandez01,whelan12} in their spectra,  and/or spectro-astrometry of FELs  \citep[][and ref. therein]{whelan05}. So far, no observation of resolved molecular hydrogen emission line (MHEL) regions has existed for VLMSs or BDs. The presence of MHEL regions are typical of the spectrum of low-mass Class I sources \citep{davis_MHEL,rebeca08,davis11}, and they are usually associated with an FEL region. Thus, one should also expect MHEL regions around young BDs and VLMSs.

In this context, we present here the first \h\,1-0S(1) spectro-imaging of an outflow from a Class I VLMS, IRS54 (YLW\,52). This source  ($\alpha$=16:27:51.7, $\delta$=-24:31:46.0) is located outside of the main clouds in the Ophiuchus star-forming region, and it has been classified as a late-stage Class I source with a bolometric luminosity of  $\sim$0.78\,\lsun\ \citep{vankempen09}.
IRS54 is thus in its main accretion phase, representing one of the lowest luminosity sources for which an \h\ outflow has been spatially resolved.

\section{Observations and data analysis}

The data were acquired on 14 June 2010 at the Very Large Telescope at Paranal Observatory, Chile, using the integral field spectrograph SINFONI at medium resolution in the K-band (R$\sim$4000). The chosen pixel scale was 100\,mas, corresponding to a  3\arcsec x3\arcsec\ field of view. The observations were acquired under 0\farcs4 seeing (DIMM FWHM), leading to a spatial resolution of 48\,AU at the location of IRS54 (d$\sim$120\,pc). The total integration time is 1200\,s.
To correct for atmospheric response, observations of a telluric standard star of spectral type B were performed.  
The main data reduction process was done using the SINFONI data-reduction pipeline, i.e., dark and bad pixel masks, flat-field corrections, optical distortion correction, and wavelength calibration using arc lamps. 
To test the goodness of the wavelength calibration, we applied the final wavelength transformation matrix to a sky cube. By measuring the wavelength of the OH lines present in the cube, we found a systematic wavelength shift of $\sim$2.2\AA\ with respect to their theoretical value. The final data cube was thus shifted in wavelength to take this error into account.
In addition to the SINFONI pipeline, the STARLINK software was used to correct the spectrum from atmospheric absorption and to flux-calibrate the data. With this aim, the standard star spectrum was extracted by collapsing the central region of the data cube. The Br$\gamma$ line was then removed from the standard spectrum before dividing it by a normalised blackbody at the appropriate temperature. Then, the spectrum was grown up into a 60x70 pixel cube, as was that of our SINFONI data, to correct for the telluric features and to flux-calibrate the data.

The brightest lines present in the K-band SINFONI spectrum are H$_2$ lines from different rovibrational levels and the \brg\ line. 
Because both continuum and line emissions are observed simultaneously, an accurate subtraction of the continuum can be performed, 
allowing us to obtain very clean images of the different line-emitting regions. To better probe the jet morphology close to the central source, the continuum emission was removed from the IFS cube using the IRAF subroutine ``CONTINUUM'' iteratively along the dispersion axis at each spatial position of the datacube. All images shown in this letter have been continuum-subtracted using this method.
In addition, integral field spectroscopy enables us to obtain precise positional measurements of emission knots relative to each other and to the source continuum. For instance, a $\sim$16\,mas positional accuracy can be obtained by taking our seeing conditions and a signal-to-noise ratio of only $\sim$10 into account (position accuracy $\sim$Seeing/(2.35$\times$SN); \citealt{davis11}).
 
%\section{Stellar properties}

\section{Outflow physics and morphology}

Integral field spectroscopic observations allow us to retrieve direct information about the morphology and kinematics of the H$_2$ emission close to the Class I protostar IRS54. 
Indeed, the \h\,2.122\,\um\ line is one of the brightest lines tracing outflow activity in the K-band. 
%Thanks to the SINFONI IFU observations we can retrieve direct information about the morphology and kinematics of this line in the region close to the Class I protostar IRS54. This source is located at only $\sim$120\,pc.  

\begin{figure}[!th]
\centering
\resizebox{0.7\columnwidth}{!}{\includegraphics{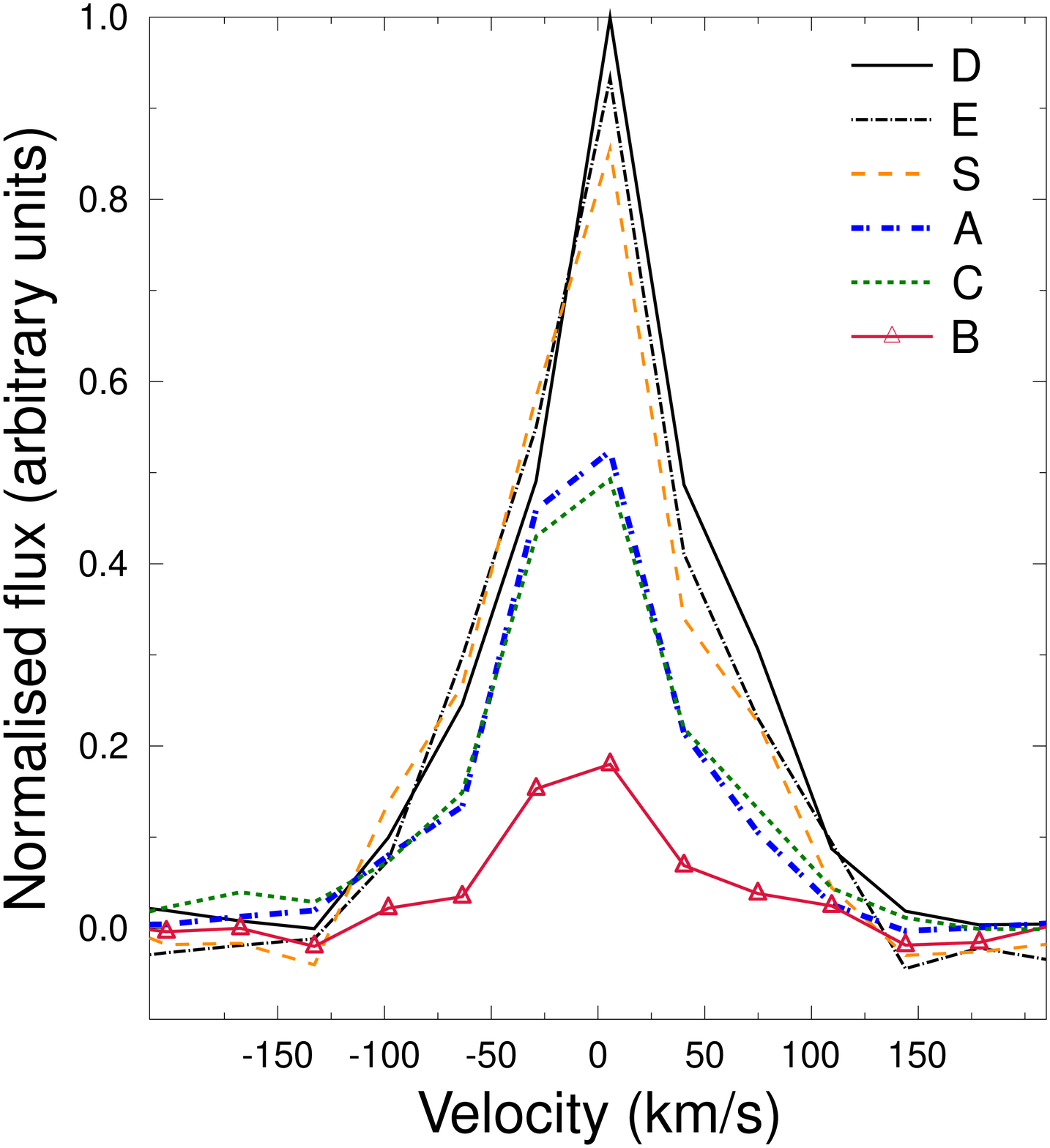}}
\caption{Line profile of the H$_2$\,2.122\,\um\ line extracted for six different spatial positions (see Fig.\,\ref{fig:vel_channel} and Table\,\ref{tab:parameters} for information about the spatial distribution of the different extracted areas). All line profiles are normalised to the flux of knot D.}
\label{fig:line_profile}
\end{figure}

Figure\,\ref{fig:vel_channel} shows the averaged continuum-subtracted \h\,1-0S(1) emission across the datacube (panel \textit{d}), along with the blue- and red-shifted \h\ outflow components (panels \textit{a} and \textit{c}). In addition, the \h\ emission at rest velocity (corresponding to one pixel in the datacube dispersion direction) is shown in panel \textit{b}. The \h\,1-0S(1) spectral image in panel \textit{d} was  contructed by averaging five different velocity channels at the local standard of rest (from $-64$\,\kms to +75\,\kms), while the blue- and red-shifted images correspond to the averaged emission over two pixels in the dispersion direction (-64\,\kms\ and -30\,\kms, and at +40\,\kms\ and +75\,\kms).
The figure shows a very complex H$_2$ morphology with gas displaying an X-shaped spatial distribution superimposed on a more collimated structure, possibly jet-like (see discussion bellow), located westward of IRS54 and extending over $\sim$1\arcsec\ ($\sim$120\,AU). All regions show a knotty structure, with condensations A, B, C, E, and D displaced from the source ($\Delta \alpha$,$\Delta \delta$) at about (1\farcs3,-0\farcs4), (0\farcs4, 0\farcs5), (0\farcs3, 0\farcs4), (-0\farcs1,0), and (-0\farcs3,0), respectively (see panel \textit{d}). The overall gas distribution is very similar to the one observed in molecular jets from embedded low-mass protostars, as usually traced by CO/molecular observations (e.g., \citealt{tafalla04}). In this case, however, two well-defined red- and blue-shifted lobes cannot be identified. As shown from the velocity channels represented from panels \textit{a} to \textit{d}, red- and blue-shifted emission can be roughly associated with the same spatial regions.
It should be noted that the velocities of the channel maps do not correspond to the \h\,2.122\,\um\ line peak velocity at any spatial position. Indeed, this line shows a very broad profile with a full width zero intensity of $\sim$200\,\kms, but with a peak velocity of $\sim$0\,\kms\ (see Fig.\,\ref{fig:line_profile}). This may suggest that the outflow lies very close to the plane of the sky.
%%%%%%%%%%%%%%%%%%%%%%%%%%%%%%%%%%%%%%%%%%%%%%%%%%%%%%%%%%%%%%%%%%%%%%%%%%%%%%%%%%%%%%%%%%%%%%%%%%%%%%%%%%%%%%%
%%%%%%%%%%%%%%%%%%%%%%%%%%%%%%%%%%%%%%%%%%%%%%%%%%%%%%%%%%%%%%%%%%%%%%%%%%%%%%%%%%%%%%%%%%%%%%%%%%%%%%%%%%%%%%%%

In addition, thanks to detecting several \h\ lines, excitation diagrams could be constructed  to derive the temperature, \h\ column density, and extinction of the molecular gas \citep[e.g.,][]{ale06,rebeca10}. The results are shown in Table\,\ref{tab:parameters}. 
The \h\ line fluxes (see Table\,\ref{tab:fluxes}) were measured at each of the positions indicated in Fig.\,\ref{fig:vel_channel} by simulating ``long-slit'' spectra with a fixed slit width corresponding to an extraction aperture of 3x3 pixels.
Temperature and N(\h) values range from $\sim$1840\,K to 3340\,K and from 2.4 to 20$\times$10$^{16}$\,cm$^{-2}$. The visual extinction, temperature, and N(\h) are higher westward of IRS54, increasing along the jet-like structure towards the source position (knot D and on source), which suggests
that the western lobe is the red-shifted lobe of the outflow and that the X-shaped and jet-like structures have different physical conditions. The extinction values found towards the source are very similar to those reported in previous works, where A$_V$ values $\gtrsim$25\,mag are reported \citep{haisch04}.
%The highest values of the N(\h) and extinction westward of IRAS54 may indicate that emission as the red lobe of the outflow. 

Finally, we note that the \brg\ emission is not spatially resolved, and it matches the stellar continuum.

\section{Accretion and ejection properties}

From our observations, it is also possible to derive the accretion luminosity and the mass accretion and ejection rates of IRS54. 
Using the relation between the luminosity of the \brg\ line (L(\brg)) and the accretion luminosity (\lacc) derived by \cite{calvet04}, we found an accretion luminosity of \lacc$\sim$0.64\,\lsun. The luminosity of the \brg\ line was measured from the integrated flux across the \brg\ line in our cube (F=4.08$\times$10$^{-14}$\,\ergscm) and corrected by a visual extinction of 30\,mag. 
Considering the modest bolometric luminosity of this source (\lbol$\sim$0.78\lsun; \citealt{vankempen09}), the accretion luminosity is very high, with an \lacc/\lbol value of $\sim$80\%, which is consistent with a very young and active YSO. This supports the idea that IRS54 is a young YSO \citep{vankempen09} still accreting large amounts of circumstellar matter. 
Assuming L$_{bol}$=\lacc+L$_*$, we find a stellar luminosity of L$_*\sim$0.14\,\lsun. We derive a stellar mass and radius of 0.1-0.2\,M$_\odot$ and $\sim$2\,R$_\odot$ by placing the stellar luminosity into an HR diagram and considering the bithline locus using the evolutionary tracks of \cite{siess00} and \cite{stahler88}.
This indicates that IRS54 has spectral type M and is a very low-mass star.

A mass accretion rate of $\sim$3.0$\times$10$^{-7}$\,\msyr\ is inferred from the observed \lacc, following the expression \macc=(L$_{acc}$\,R$_*$\,/\,G\,M$_*$)$\times$(1-R$_*$/R$_i$), R$_i$ being the inner radius of the accretion disk\,\citep[with R$_i$=5\,R$_*$, see][]{gullbring98}. The derived value is higher than those found in more evolved objects of roughly the same mass \citep{natta06, gatti06}, again pointing  to the young nature of this source.

%%%%%%%%%%%%%%%%%%%%	H2 diagnostics	%%%%%%%%%%%%%%%%%%%%%%%%%%%%%%%%%%%%%
\begin{table}
\begin{minipage}[t]{\columnwidth}
\caption{H$_2$ diagnostics.}
\label{tab:parameters}
\centering
\renewcommand{\footnoterule}{}  % to avoid a line before footnotes
\begin{tabular}{c c c c c}
\hline \hline
knot & d & A$_V$ & T & N(H$_2$)	\\
     & (\arcsec) & (mag) & (K) & 10$^{16}$\,cm$^{-2}$  \\ 
\hline
source  & [0,0]	    & 30	& 3342$\pm$250	& 20$_{3.8}^{+3.4}$	\\
A       & [1.3,-0.4] & 20	& 2001$\pm$77	& 3.2$_{-0.2}^{+0.2}$	\\
B       & [0.4,0.55]	& 15	& 2008$\pm$42	& 2.4$_{-0.1}^{+0.1}$  	\\
C       & [0.3,0.4]	& 17	& 1839$\pm$59	& 3.8$_{-0.2}^{+0.2}$	\\
D       &  [-0.3,0] 	& 25	& 2009$\pm$55	& 14.5$_{-0.7}^{+0.8}$	\\
\hline
\end{tabular}
\end{minipage}
\end{table}
%%%%%%%%%%%%%%%%%%%%%%%%%%%%%%%%%%%%%%%%%%%%%%%%%%%%%%%%%%%%%%%%%%%%%%%%%%%%

The mass-loss rate carried by the warm molecular component ($\dot{M}_{H_2}$) can be computed from the derived N(\h) values using the expression \mh=2\,$\mu$\,m$_H$N(H$_2$)\,A\,dv$_t$/dl$_t$ (see, e.g., \citealt{davis_MHEL,davis11}). 
Here, A is the area of the emitting region, $\mu$ is the mean atomic weight, and dl$_t$ and dv$_t$ are the projected length and the tangential velocity.  We have assumed that the area where the molecular hydrogen is emitted is equal to the extent of the flow along the jet axis from position S to D ($\sim$0\farcs5) multiplied by the width of the flow (i.e. the seeing). We have considered an average N(\h) value for this region of  $\sim$1.7$\times$10$^{17}$\,cm$^{-2}$. 
To retrieve v$_t$, the inclination angle of the outflow with respect to the line of sight should be known. 
As outlined before, the radial velocities suggest that the outflow is almost in the plane of the sky, and thus, an estimate of the jet velocity from an unknown flow angle is too uncertain. Therefore, we assume the width of the \h\ lines presented in Fig.\,\ref{fig:line_profile} as a lower limit to the jet velocity, i.e. $\sim$200\,\kms. 
%Therefore, we assume an angle between 5\degr\ and 30\degr\ and an average radial velocity of the flow of $\sim$7\,\kms. 
This gives a value of \mh$\gtrsim$1.6$\times$10$^{-10}$\,\msyr\ which is about two orders of magnitude lower than expected if compared with the derived \macc\ value ($\dot{M}_{out}$/\macc$\sim$0.1).
The computed \mh\ value is also around two to three orders of magnitude lower than the \mh\ value derived for low-mass Class I sources using the same technique \citep{davis11}. However, it is worth noting that these sources are likely to be more massive than ours.
This suggests that (1) most of the outflow material is transported by a cooler and denser component than traced by the near-IR \h\ lines, such as the \h\ pure rotational lines \citep{ale08}; or (2) most of the \h\ has been dissociated, and the jet is mainly atomic. The latter has been observed in other Class I jets, where the measured \mh\ is around one order of magnitude lower than measured from the \FeII\ emission \citep{davis11,nisini_hh1}.

%%%%%%%%%%%%%%%%%%%%%%%%%%%%%%%%%%%%%%%%%%%%%%%%%%%%%%%%%%%%%%%%%%%%%%%%%%%%%%%%%%%%%%%%%%%%%%%%%%%%%%%%%%%%%%%
%\begin{figure}[th]
%\centering
%\resizebox{0.8\columnwidth}{!}{\includegraphics{sketch+image3.eps}}
%\caption{Sketch showing IRS54 outflow/jet morphology.}
%\label{fig:sketch}
%\end{figure}
%%%%%%%%%%%%%%%%%%%%%%%%%%%%%%%%%%%%%%%%%%%%%%%%%%%%%%%%%%%%%%%%%%%%%%%%%%%%%%%%%%%%%%%%%%%%%%%%%%%%%%%%%%%%%%%%

\section{Discussion}

Molecular outflows, \h\ jets in particular, have been observed from Class 0 to Class II sources through a wide range of masses. Here, we show that MHEL regions are also common in the first evolutionary stages of VLMSs, suggesting that the same launching mechanism is acting from very low- to high-mass stars and across a wide range of evolutionary stages.

In general, MHEL properties are consistent with shock-heated gas from the inner regions of  Herbig-Haro objects or from spatially extended wide-angle winds \citep{davis_MHEL, ale06,beck08, davis11}. 
The reported gas temperatures and excitation properties of IRS54 are consistent with shock-heated material, as well. Similar values have also been found in low-mass Class I sources and CTTSs \citep{beck08,davis11}.
The overall outflow structure can be interpreted as the sum of \h\ emission excited along a wide-angle cavity, plus the contribution from a jet-like structure. This latter structure is, however, only observed in one of the lobes. This probably indicates different excitation conditions for the two lobes of the jet, i.e. an asymmetric jet in which the blue-shifted lobe might have a higher velocity, dissociating the molecular \h\ component. 
If this is the case, ionic/atomic emission should be present in the eastern jet.
Velocity asymmetries have been observed before in other protostellar jets, such as RW\,Aur \citep[e.g.,][]{melnikov09}. 

%With respect to the kinematics, IRS54 shows velocity channel maps very similar to those typically found in low and high-mass outflows: at low velocities, the outflow seems to trace a wide-angle cavity, whilst at higher velocities, the outflow appears to be jet-like.
%This latter structure is, however, only observed in one of the lobes. This likely indicates different excitation conditions for the two lobes of the jet, i.e. an asymmetric jet in which the blue-shifted lobe might have a higher velocity, dissociating the molecular \h\ component. 
%If this is the case, ionic/atomic emission should be present in the eastern jet.
%Velocity asymmetries have been observed before in other protostellar jets, such as RW\,Aur \citep[e.g.,][]{melnikov09}.

IRS 54 is known to possess a large-scale S-shaped bipolar jet as traced by a chain of \h\ knots\footnote{named MHO\,2128-2131, see http://www.astro.ljmu.ac.uk/MHCat/} extending  over several arcminutes \citep[feature f08-01 in][]{tigran04}. 
%Further observations involving atomic species are needed in order to trace high excitation conditions and confirm this hypothesis.
%Wide-field \h\ images indicates IRS54 as the possible excitation source of a bipolar jet outlined by a chain of \h\ knots extending over several arcminutes \citep{tigran04,jorgensen09}. 
From wide-field \h\ images, the jet appears to precess with a position angle (PA) changing from $\sim$72\degr\ in the outermost knots passing through $\sim$78\degr\ in the middle part of the jet down to 90\degr\ closest to the source (MHO\,2129; see also Fig.\,B.1 in \citealt{jorgensen09}). The strong jet precession may then be at the origin of the outflow cavity, suggesting that the low-velocity \h\ emission is gas excited along the cavity walls by the interaction of a wide-angle wind with the ambient material. On the other hand, the PA of the internal jet-like structure detected in our \h\ spectral images  is consistent with the PA of the innermost knots of the large-scale \h\ jet (i.e., MHO\,2129). This may support the idea that the denser and more reddened jet-like structure in our images is indeed the upstream region of the \h\ jet observed in the wide-field images.
Jet precession alone can, however, hardly account for all the observed features shown in Fig.\,\ref{fig:vel_channel}. 
By assuming a lower limit to the jet velocity of $\sim$200\,\kms, the dynamical age of condensations A, B, C, D, and E are around 3.9, 1.9, 1.4, 0.9, and 0.3\,yr. This would lead to a precession period of about four years in order to generate the structures (A, B, and C) at each side of the so-called cavity (opening angle $\sim$65\degr). This period is too short when compared with typical jet fast precession periods \citep{rosen04,ale08}, and not consistent with the longer period suggested by the large-scale jet images \citep[see f08-01 field in][]{tigran04}.

On the other hand, a wide-angled wind might be able to explain both the broad line profiles presented in Fig.\,\ref{fig:line_profile} and the observed features in Fig.\,\ref{fig:vel_channel}.   
Considering the jet-like structure alone (condensations E and D), we have measured the \h\ jet width as a function of the distance from the source (see, Fig.\,\ref{fig:width}). We estimate a full opening angle of the flow of $\sim$23\,\degr, fitting a straight line to the points from 40 to 140\,AU in Fig.\,\ref{fig:width} \citep[see][for more details]{davis11}. Similar values are also found in jets from low-mass Class I sources and CTTSs, which show opening angles between 20\degr\ and 42\degr\ \citep{ hartigan04,davis11}. Beyond $\gtrsim$50\,AU, the jet width slowly increases with distance \citep[see Fig.\,\ref{fig:width} and][]{hartigan04, davis11} as expected for a free lateral expansion of a supersonic jet \citep{cabrit07_lecture}. Therefore, the jet collimation must take place within the first $\lesssim$50\,AU from the source, in agreement with MHD wind models \citep{dougados04}.
\cite{despina12} show that molecular hydrogen can survive along MHD disk-wind streamlines. The presence of \h\ emission detected at only $\sim$48\,AU from the source, together with the broad line profiles shown in Fig.\,\ref{fig:line_profile}, tend to favour the presence of an MHD disk-wind model.
Nevertheless, a wide-angled wind alone cannot account for the precessing jet observed in the large-scale images.
Therefore, a combination of both a wide-angled wind and a precessing jet is possibly the main mechanism behind the complex \h\ structure in IRS54 (Fig.\,\ref{fig:sketch}).

\section{Conclusions}
In this letter, we present the first spatially resolved \h\ emission (MHEL) region around IRS54, a Class I VLMS. The \h\ emission was detected down to the first $\sim$50\,AU from the source, and it shows a very complex morphology. The emission might be interpreted as coming from the interaction of a wide-angle wind with an outflow cavity and a molecular jet. In addition, the detection of several \h\ line transitions and the \brg\ line allows us to derive various accretion/ejection properties:
\begin{itemize}
 \item We computed the extinction, \h\ column density, and temperature values at different outflow spatial positions from the analysis of excitation diagrams. The highest values are found westward of IRS54, increasing along the jet-like structure towards the source position. 
Average values of A$_v\sim$28\,mag, T=2000-3000\,K, and N(\h)$\sim$1.7$\times$10$^{17}$\,cm$^{-2}$ are found along the jet-like structure (knots D and S).
 
 \item We inferred an accretion luminosity and mass accretion rate of 0.64\,\lsun\ and 3$\times$10$^{-7}$\,\msyr\ from the total flux of the \brg\ emission. From the computed \lacc\ and the \lbol\ value found in literature, we derived L$_*\sim$0.14\,\lsun, which yields a stellar mass of $\sim$0.1-0.2\,\msun.The accretion luminosity accounts for $\sim$80\% of the total luminosity. This, together with the high \macc\ value, points to the young nature of IRS54.

\end{itemize}

\begin{acknowledgements}
We thank the anonymous referee for the comments that helped improve the paper. 
\end{acknowledgements}

\bibliographystyle{aa}
\bibliography{references}

\begin{thebibliography}{34}
\expandafter\ifx\csname natexlab\endcsname\relax\def\natexlab#1{#1}\fi

\bibitem[{{Andr{\'e}} {et~al.}(2007){Andr{\'e}}, {Belloche}, {Motte}, \&
  {Peretto}}]{andre07}
{Andr{\'e}}, P., {Belloche}, A., {Motte}, F., \& {Peretto}, N. 2007, \aap, 472,
  519

\bibitem[{{Arce} {et~al.}(2007){Arce}, {Shepherd}, {Gueth}, {Lee}, {Bachiller},
  {Rosen}, \& {Beuther}}]{arce07}
{Arce}, H.~G., {Shepherd}, D., {Gueth}, F., {et~al.} 2007, Protostars and
  Planets V, 245

\bibitem[{{Bachiller} \& {Tafalla}(1999)}]{bachiller99}
{Bachiller}, R. \& {Tafalla}, M. 1999, in NATO ASIC Proc. 540: The Origin of
  Stars and Planetary Systems, ed. C.~J. {Lada} \& N.~D. {Kylafis}, 227

\bibitem[{{Beck} {et~al.}(2008){Beck}, {McGregor}, {Takami}, \& {Pyo}}]{beck08}
{Beck}, T.~L., {McGregor}, P.~J., {Takami}, M., \& {Pyo}, T.-S. 2008, \apj,
  676, 472

\bibitem[{{Cabrit}(2007)}]{cabrit07_lecture}
{Cabrit}, S. 2007, in Lecture Notes in Physics, Berlin Springer Verlag, Vol.
  723, Lecture Notes in Physics, Berlin Springer Verlag, ed. J.~{Ferreira},
  C.~{Dougados}, \& E.~{Whelan}, 21

\bibitem[{{Calvet} {et~al.}(2004){Calvet}, {Muzerolle}, {Brice{\~n}o},
  {Hern{\'a}ndez}, {Hartmann}, {Saucedo}, \& {Gordon}}]{calvet04}
{Calvet}, N., {Muzerolle}, J., {Brice{\~n}o}, C., {et~al.} 2004, \aj, 128, 1294

\bibitem[{{Caratti o Garatti} {et~al.}(2008){Caratti o Garatti}, {Froebrich},
  {Eisl{\"o}ffel}, {Giannini}, \& {Nisini}}]{ale08}
{Caratti o Garatti}, A., {Froebrich}, D., {Eisl{\"o}ffel}, J., {Giannini}, T.,
  \& {Nisini}, B. 2008, \aap, 485, 137

\bibitem[{{Caratti o Garatti} {et~al.}(2006){Caratti o Garatti}, {Giannini},
  {Nisini}, \& {Lorenzetti}}]{ale06}
{Caratti o Garatti}, A., {Giannini}, T., {Nisini}, B., \& {Lorenzetti}, D.
  2006, \aap, 449, 1077

\bibitem[{{Davis} {et~al.}(2011){Davis}, {Cervantes}, {Nisini}, {Giannini},
  {Takami}, {Whelan}, {Smith}, {Ray}, {Chrysostomou}, \& {Pyo}}]{davis11}
{Davis}, C.~J., {Cervantes}, B., {Nisini}, B., {et~al.} 2011, \aap, 528, A3

\bibitem[{{Davis} {et~al.}(2001){Davis}, {Ray}, {Desroches}, \&
  {Aspin}}]{davis_MHEL}
{Davis}, C.~J., {Ray}, T.~P., {Desroches}, L., \& {Aspin}, C. 2001, \mnras,
  326, 524

\bibitem[{{Dougados} {et~al.}(2004){Dougados}, {Cabrit}, {Ferreira}, {Pesenti},
  {Garcia}, \& {O'Brien}}]{dougados04}
{Dougados}, C., {Cabrit}, S., {Ferreira}, J., {et~al.} 2004, \apss, 292, 643

\bibitem[{{Fern{\'a}ndez} \& {Comer{\'o}n}(2001)}]{fernandez01}
{Fern{\'a}ndez}, M. \& {Comer{\'o}n}, F. 2001, \aap, 380, 264

\bibitem[{{Garcia Lopez} {et~al.}(2010){Garcia Lopez}, {Nisini},
  {Eisl{\"o}ffel}, {Giannini}, {Bacciotti}, \& {Podio}}]{rebeca10}
{Garcia Lopez}, R., {Nisini}, B., {Eisl{\"o}ffel}, J., {et~al.} 2010, \aap,
  511, A5

\bibitem[{{Garcia Lopez} {et~al.}(2008){Garcia Lopez}, {Nisini}, {Giannini},
  {Eisl{\"o}ffel}, {Bacciotti}, \& {Podio}}]{rebeca08}
{Garcia Lopez}, R., {Nisini}, B., {Giannini}, T., {et~al.} 2008, \aap, 487,
  1019

\bibitem[{{Gatti} {et~al.}(2006){Gatti}, {Testi}, {Natta}, {Randich}, \&
  {Muzerolle}}]{gatti06}
{Gatti}, T., {Testi}, L., {Natta}, A., {Randich}, S., \& {Muzerolle}, J. 2006,
  \aap, 460, 547

\bibitem[{{Gullbring} {et~al.}(1998){Gullbring}, {Hartmann}, {Briceno}, \&
  {Calvet}}]{gullbring98}
{Gullbring}, E., {Hartmann}, L., {Briceno}, C., \& {Calvet}, N. 1998, \apj,
  492, 323

\bibitem[{{Haisch} {et~al.}(2004){Haisch}, {Greene}, {Barsony}, \&
  {Stahler}}]{haisch04}
{Haisch}, Jr., K.~E., {Greene}, T.~P., {Barsony}, M., \& {Stahler}, S.~W. 2004,
  \aj, 127, 1747

\bibitem[{{Hartigan} {et~al.}(2004){Hartigan}, {Edwards}, \&
  {Pierson}}]{hartigan04}
{Hartigan}, P., {Edwards}, S., \& {Pierson}, R. 2004, \apj, 609, 261

\bibitem[{{J{\o}rgensen} {et~al.}(2009){J{\o}rgensen}, {van Dishoeck},
  {Visser}, {Bourke}, {Wilner}, {Lommen}, {Hogerheijde}, \&
  {Myers}}]{jorgensen09}
{J{\o}rgensen}, J.~K., {van Dishoeck}, E.~F., {Visser}, R., {et~al.} 2009,
  \aap, 507, 861

\bibitem[{{Khanzadyan} {et~al.}(2004){Khanzadyan}, {Gredel}, {Smith}, \&
  {Stanke}}]{tigran04}
{Khanzadyan}, T., {Gredel}, R., {Smith}, M.~D., \& {Stanke}, T. 2004, \aap,
  426, 171

\bibitem[{{Melnikov} {et~al.}(2009){Melnikov}, {Eisl{\"o}ffel}, {Bacciotti},
  {Woitas}, \& {Ray}}]{melnikov09}
{Melnikov}, S.~Y., {Eisl{\"o}ffel}, J., {Bacciotti}, F., {Woitas}, J., \&
  {Ray}, T.~P. 2009, \aap, 506, 763

\bibitem[{{Natta} {et~al.}(2006){Natta}, {Testi}, \& {Randich}}]{natta06}
{Natta}, A., {Testi}, L., \& {Randich}, S. 2006, \aap, 452, 245

\bibitem[{{Nisini} {et~al.}(2005){Nisini}, {Bacciotti}, {Giannini}, {Massi},
  {Eisl{\"o}ffel}, {Podio}, \& {Ray}}]{nisini_hh1}
{Nisini}, B., {Bacciotti}, F., {Giannini}, T., {et~al.} 2005, \aap, 441, 159
  (N05)

\bibitem[{{Panoglou} {et~al.}(2012){Panoglou}, {Cabrit}, {Pineau Des
  For{\^e}ts}, {Garcia}, {Ferreira}, \& {Casse}}]{despina12}
{Panoglou}, D., {Cabrit}, S., {Pineau Des For{\^e}ts}, G., {et~al.} 2012, \aap,
  538, A2

\bibitem[{{Phan-Bao} {et~al.}(2011){Phan-Bao}, {Lee}, {Ho}, \& {Tang}}]{phan11}
{Phan-Bao}, N., {Lee}, C.-F., {Ho}, P.~T.~P., \& {Tang}, Y.-W. 2011, \apj, 735,
  14

\bibitem[{{Phan-Bao} {et~al.}(2008){Phan-Bao}, {Riaz}, {Lee}, {Tang}, {Ho},
  {Mart{\'{\i}}n}, {Lim}, {Ohashi}, \& {Shang}}]{phan08}
{Phan-Bao}, N., {Riaz}, B., {Lee}, C.-F., {et~al.} 2008, \apjl, 689, L141

\bibitem[{{Rosen} \& {Smith}(2004)}]{rosen04}
{Rosen}, A. \& {Smith}, M.~D. 2004, \mnras, 347, 1097

\bibitem[{{Siess} {et~al.}(2000){Siess}, {Dufour}, \& {Forestini}}]{siess00}
{Siess}, L., {Dufour}, E., \& {Forestini}, M. 2000, \aap, 358, 593

\bibitem[{{Stahler}(1988)}]{stahler88}
{Stahler}, S.~W. 1988, \apj, 332, 804

\bibitem[{{Tafalla} {et~al.}(2004){Tafalla}, {Santiago}, {Johnstone}, \&
  {Bachiller}}]{tafalla04}
{Tafalla}, M., {Santiago}, J., {Johnstone}, D., \& {Bachiller}, R. 2004, \aap,
  423, L21

\bibitem[{{van Kempen} {et~al.}(2009){van Kempen}, {van Dishoeck},
  {Hogerheijde}, \& {G{\"u}sten}}]{vankempen09}
{van Kempen}, T.~A., {van Dishoeck}, E.~F., {Hogerheijde}, M.~R., \&
  {G{\"u}sten}, R. 2009, \aap, 508, 259

\bibitem[{{Whelan} {et~al.}(2012){Whelan}, {Ray}, {Comeron}, {Bacciotti}, \&
  {Kavanagh}}]{whelan12}
{Whelan}, E., {Ray}, T., {Comeron}, F., {Bacciotti}, F., \& {Kavanagh}, P.
  2012, ArXiv e-prints

\bibitem[{{Whelan} {et~al.}(2005){Whelan}, {Ray}, {Bacciotti}, {Natta},
  {Testi}, \& {Randich}}]{whelan05}
{Whelan}, E.~T., {Ray}, T.~P., {Bacciotti}, F., {et~al.} 2005, \nat, 435, 652

\bibitem[{{Wouterloot} {et~al.}(2005){Wouterloot}, {Brand}, \&
  {Henkel}}]{wouterloot05}
{Wouterloot}, J.~G.~A., {Brand}, J., \& {Henkel}, C. 2005, \aap, 430, 549

\end{thebibliography}

\Online

\begin{appendix}

\onecolumn

\section{\h\ line fluxes and spectra}

\begin{figure*}[th]
\centering
\includegraphics[width=16.4cm,clip]{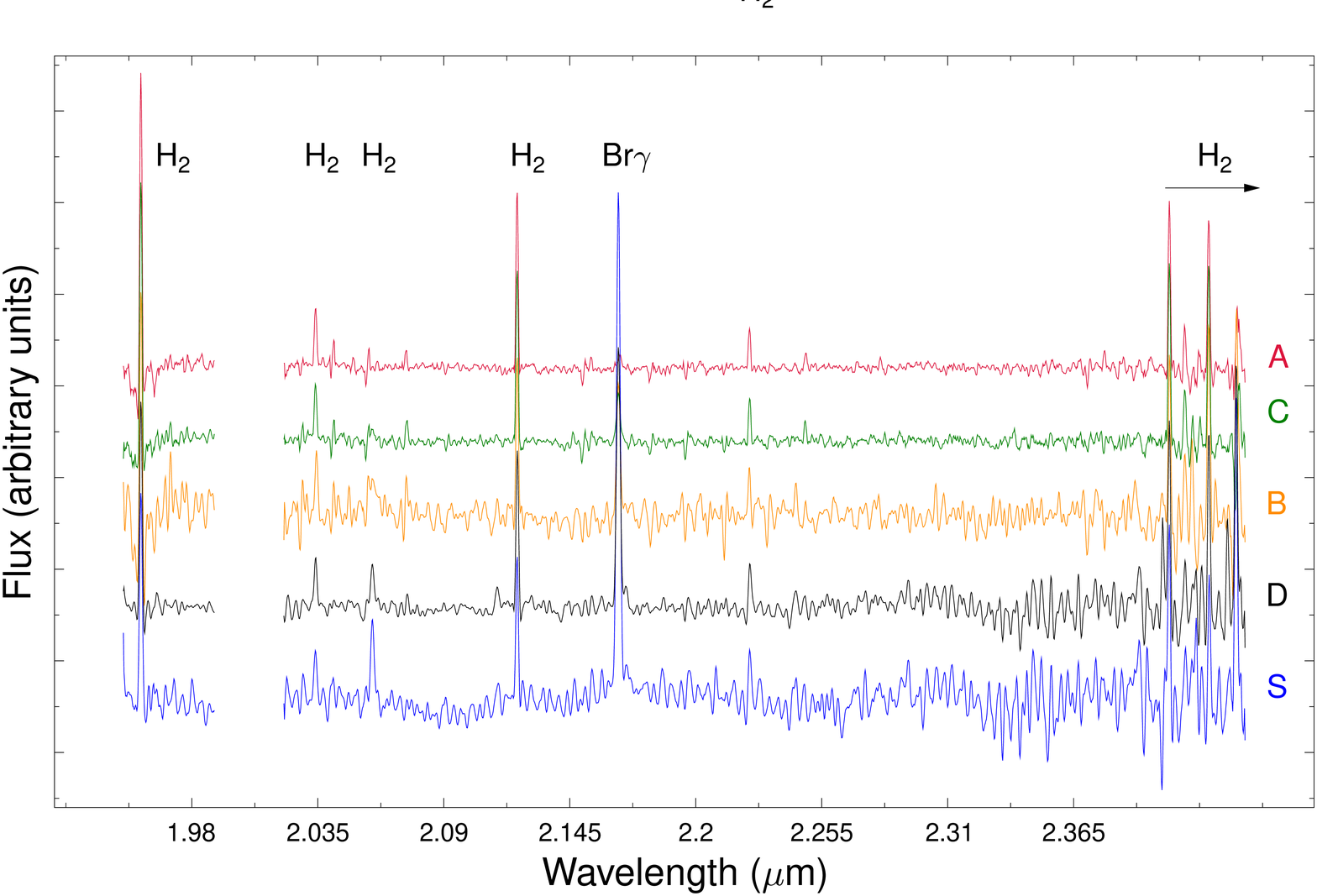}
\caption{Continuum-subtracted spectra at five different positions along the outflow (see Fig.\,\ref{fig:vel_channel} and Tab\,\ref{tab:fluxes}).}
\label{fig:spectra}
\end{figure*}
%%%%%%%%%%%%%%%%%%%%%%%%%%%%%%%%%%%%%%%%%%%%%%%%%%%%%%%%%%%%%%%%%%%%%%%%%%%%%%%%%%%%%%%%%%%%%%%%%%%%%%%%%%%%%%%%

%%%%%%%%%%%%%%%%%%%	H2 line fluxes	%%%%%%%%%%%%%%%%%%%%%%%%%%%%%%%%%%%%%
\begin{table*}[!bh]
\begin{minipage}[!bh]{\textwidth}
\caption{H$_2$ line fluxes.}
\label{tab:fluxes}
\centering
\renewcommand{\footnoterule}{}  % to avoid a line before footnotes
\begin{tabular}{c c c c c c c c c}
\hline \hline
knot & d & 1-0S(2) & 1-0S(1) & 1-0S(0) & 2-1S(1) & 1-0Q(1) & 1-0Q(2) & 1-0Q(3)  	\\
     & (\arcsec) &\multicolumn{7}{c}{10$^{-17}$\,erg/s/cm$^{2}$}  \\ 
\hline
S(source)  & [0,0]	    & 2.53$\pm$0.43 & 5.55$\pm$0.14 & 1.75$\pm$0.31 &               & 7.43$\pm$0.62 &               & 7.95$\pm$0.79  \\
A  & [1.3,-0.4] & 0.94$\pm$0.01 & 2.41$\pm$0.03 & 0.60$\pm$0.05 & 0.24$\pm$0.02 & 2.61$\pm$0.08 & 0.77$\pm$0.11 & 2.42$\pm$0.13 \\
B  & [0.4,0.55]	& 1.07$\pm$0.02 & 2.96$\pm$0.03 & 0.81$\pm$0.02 & 0.30$\pm$0.02 & 3.02$\pm$0.05 & 0.93$\pm$0.14 & 2.99$\pm$0.09  \\
C  & [0.3,0.4]	& 1.09$\pm$0.03 & 3.01$\pm$0.04 & 0.94$\pm$0.06 & 0.26$\pm$0.02 & 3.55$\pm$0.19 & 1.11$\pm$0.16 & 2.99$\pm$0.09  \\
D  & [-0.3,0] 	& 2.11$\pm$0.19 & 6.02$\pm$0.09 & 1.82$\pm$0.15 & 0.69$\pm$0.13 & 7.80$\pm$0.41 & 2.30$\pm$0.38 & 7.91$\pm$0.52   \\
%E  & [-0.1,0]	& 2.54$\pm$0.36 & 5.91$\pm$0.10 & 2.29$\pm$0.28 &               & 7.37	$\pm$0.70 & 1.66$\pm$0.33 & 8.29$\pm$0.67   \\

\hline
\end{tabular}
\end{minipage}
\end{table*}
%%%%%%%%%%%%%%%%%%%%%%%%%%%%%%%%%%%%%%%%%%%%%%%%%%%%%%%%%%%%%%%%%%%%%%%%%%%%

\onecolumn
\section{Outflow width and morphology}

%%%%%%%%%%%%%%%%%%%%%%%%%%%%%%%%%%%%%%%%%%%%%%%%%%%%%%%%%%%%%%%%%%%%%%%%%%%
\begin{figure*}[!h]
\sidecaption
\includegraphics[width=0.5\columnwidth]{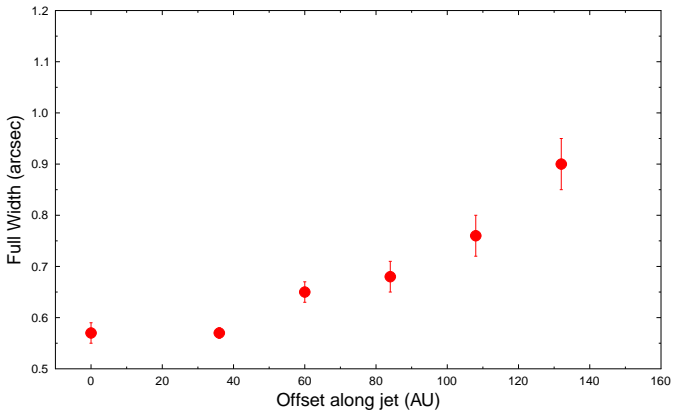}
\caption{Width of the \h\,1-0S(1) jet-like structure as a function of the distance from the source. The jet width has been estimated by extracting 2-pixel-wide vertical slices to the average \h\ spectral image presented in Fig.\,\ref{fig:vel_channel} and fitting a single Gaussian function. Zero offsets corresponds to the source positions.}
\label{fig:width}
\end{figure*}
%%%%%%%%%%%%%%%%%%%%%%%%%%%%%%%%%%%%%%%%%%%%%%%%%%%%%%%%%%%%%%%%%%%%%%%%%%%%%%%%%%%%%%%%%%%%%%%%%%%%%%%%%%%%%%%%

%%%%%%%%%%%%%%%%%%%%%%%%%%%%%%%%%%%%%%%%%%%%%%%%%%%%%%%%%%%%%%%%%%%%%%%%%%%

\begin{figure*}[!h]
\sidecaption
\includegraphics[width=0.5\columnwidth]{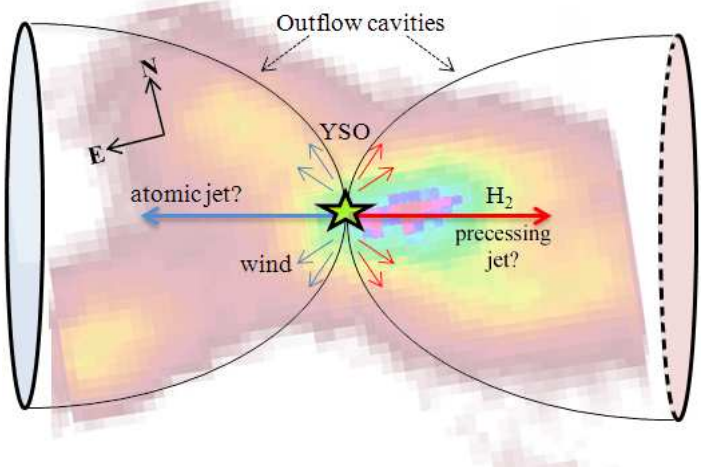}
\caption{Sketch showing IRS54 outflow/jet morphology.}
\label{fig:sketch}
\end{figure*}
%%%%%%%%%%%%%%%%%%%%%%%%%%%%%%%%%%%%%%%%%%%%%%%%%%%%%%%%%%%%%%%%%%%%%%%%%%%%%%%%%%%%%%%%%%%%%%%%%%%%%%%%%%%%%%%%

\end{appendix}

\end{document}